\documentclass[11pt,preprint,nofootinbib,showpacs,preprintnumbers,amsmath,amssymb]{revtex4}

\usepackage{exscale}
\usepackage{relsize}
\usepackage{graphicx}
\usepackage{dcolumn}
\usepackage{bm}
\usepackage{multirow}
\usepackage{CJK}
\usepackage{diagbox}
\usepackage{subfigure}
\usepackage[T1]{fontenc}
\usepackage{subfigure}

\begin{document}

\begin{CJK*}{GBK}{}

\title{Direct $CP$ violation in $D^\pm \to \pi^\pm \pi^+ \pi^-$ with $a_0^0(980)-f_0(980)$ mixing}

\author{Shi-Ji Cao$^{1,3}$\footnote{csj1999\_1215@163.com}, Jing-Juan Qi $^{2,1}$\footnote{ qijj@mail.bnu.edu.cn}, Yi-Fan Zhao$^{1}$, Chao Wang $^{3}$\footnote{chaowang@nwpu.edu.cn}, Zhen-Yang Wang$^{1}$\footnote{Corresponding author, wangzhenyang@nbu.edu.cn}, Xin-Heng Guo$^{4 }$\footnote{Corresponding author, xhguo@bnu.edu.cn}}

\affiliation{\small $^{1}$Physics Department, Ningbo University, Zhejiang 315211, China\\
	\small $^{2}$ College of Information and Intelligence Engineering, Zhejiang Wanli University, Zhejiang 315101, China\\
    \small $^{3}$ School of Life Science and Technology, Northwestern Polytechnical University, Xi'an 710072, China\\
    \small $^{4}$ School of Physics Science and Technology, Kunming University, Kunming 650214, China}

\date{\today\\}
\begin{abstract}
We investigate the direct $CP$ violation in the decay $D^\pm \to \pi^\pm \pi^+ \pi^-$ incorporating the $a_0^0(980)$-$f_0(980)$ mixing mechanism. The integrated mixing intensities $\overline  \xi_{fa}$ and $\overline  \xi_{af}$ are calculated 
 using meson masses and coupling constants extracted from various theoretical models and experimental data, yielding values of appreciable magnitude. We find that when the invariant mass of the $\pi^+\pi^-$pair lies near the 
$f_0(980)$ resonance, this isospin breaking mechanism can enhance the $CP$ asymmetry. The enhancement is particularly pronounced when 
$f_0(980)$ carries a significant $n\bar{n}$ quark component and the  $f_0(980)$ and $\sigma(600)$ mixing angle is approximately $26^\circ$. It is emphasized that the $a_0^0(980)$-$f_0(980)$ mixing mechanism can be taken into account in both theoretical and experimental studies of $CP$ violation in $B$ or $D$ meson decays.
\end{abstract}
\pacs{***************}
\maketitle
\end{CJK*}

\section{Introduction}
Charge-Parity ($CP$) violation is a fundamental phenomenon in particle physics with direct implications for the observed matter-antimatter asymmetry of the Universe. First established in neutral kaon decays in 1964 \cite{Christenson:1964fg}, it remains a central subject of investigation within and beyond the Standard Model (SM). In the SM, $CP$ violation originates from the irreducible complex phase of the Cabibbo-Kobayashi-Maskawa (CKM) matrix \cite{Cabibbo:1963yz,Schrenk:2005si}, which governs quark-flavor mixing via the weak interaction.

In recent years, the LHCb collaboration observed larger localized $CP$ asymmetries in the Dalitz plot of the $B$ three-body decays, notably in $B^\pm \to \pi^\pm \pi^+ \pi^-$ and $B^\pm \to K^\pm \pi^+ \pi^-$ \cite{LHCb:2022fpg}. These asymmetries can be attributed to interference between decay amplitudes from nearby resonances of different spin \cite{Zhang:2013oqa}.  Meanwhile, isospin-breaking effects, such as $\rho-\omega$ mixing and $a_0^0(980)-f_0(980)$ mixing \cite{Guo:1998eg,Wang:2016yrm,Lu:2014oba,Wang:2019tqt,Qi:2023pls}, also constitute important mechanisms for exploring $CP$ violation.  $CP$ violation effects in $D$ mesons are small, the SM provides clear predictions for them. By precisely measuring $CP$ violation in $D$ mesons, we can test the accuracy of the SM, explore new physics, and gain a deeper understanding of the non-perturbative effects of strong interactions. The LHCb collaboration measured $CP$ violation in the three-body decay $D^\pm \to \pi^\pm \pi^+ \pi^-$ \cite{LHCb:2013qzm},  no localized or overall $CP$ asymmetries were found with high-statistics data. The BESIII collaboration also analyzed the amplitude of $D^+ \to \pi^+ \pi^0 \pi^0$ and measured $CP$  asymmetries, and the results showed that no evidence for $CP$ violation is observed \cite{BESIII:2025les}. As demonstrated in the aforementioned three-body decay channels of $B$ mesons, localized $CP$ asymmetries in phase space can be enhanced by the interference different intermediate resonances, and the same mechanism should also apply to $D$ meson decays. Interference between differentresonances, such as $\rho^0(770)$ and $f_0(500)$, has already been studied in three-body decays of $D$ mesons \cite{Zhang:2015xea}, while the scalar type isospin breaking effect has not yet been investigated. Drawing inspiration from these, we will investigate the effects of the $a_0^0(980)$-$f_0(980)$ mixing mechanism on $D$ meson decays. The theoretical proposal of the $a_0^0(980)$-$f_0(980)$ mixing effect can be traced to the late 1970s \cite{Achasov:1979xc}. This isospin breaking effect results in an 8 MeV mass difference between the charged and neutral kaon thresholds when $a_0^0(980)$ and $f_0(980)$ decay into $K\bar{K}$. Over the years, the $a_0^0(980)$-$f_0(980)$ mixing has been thoroughly investigated across various processes and from multiple perspectives \cite{Colley:1967zz,Achasov:2003se,Kerbikov:2000pu,Krehl:1996rk,Close:2001ay,Kudryavtsev:2002uu,Grishina:2001zj,Black:2002ek,Wu:2010za,Cheng:2020qzc,Aliev:2018bln,Wang:2016wpc,
Sekihara:2014qxa,Liang:2019jtr,Buescher:2003cj,Amsler:2004ps,Hanhart:2003pg,Wang:2004ipa,Roca:2012cv,Close:2015rza,Dorofeev:2011zz}. The first experimental observation of this effect was made by the BESIII collaboration in the decays $J/\psi \rightarrow \phi f_0(980) \rightarrow \phi a_0^0(980) \rightarrow \phi \eta \pi^0$ and $\chi_{c1} \rightarrow a_0^0(980) \pi^0 \rightarrow f_0(980) \pi^0 \rightarrow \pi^+\pi^-\pi^0$ \cite{BESIII:2018ozj}. Similarly to $B$ decay, we expect that $a_0^0(980)$-$f_0(980)$ mixing could lead to magnified $CP$ violation in the $D^\pm \to \pi^\pm \pi^+ \pi^-$ decay. 

A study of the direct $CP$ violation in the decay $D^\pm \to \pi^\pm \pi^+ \pi^-$ with $a_0^0(980)-f_0(980)$ mixing is presented within the framework of the naive factorization approach. The organization of this paper is as follows. A brief overview of the $a_0^0(980)-f_0(980)$ mixing mechanism is given in Sect. \ref{af mix}. The formalism for the decay amplitudes and the $CP$ violation calculation is developed in Sect. \ref{Amp Acp}. Numerical results are presented in Sect. \ref{Num res}, and a summary with discussion is provided in Sect. \ref{Sum Dis}. The theoretical input parameters used in this work are summarized in Appendix \ref{A}.

\section{$a_0^0(980)-f_0(980)$ MIXING MECHANISM}\label{af mix}

\subsection{$a_0^0(980)-f_0(980)$ mixing amplitude}
When the $a_0^0(980)-f_0(980)$ mixing is active, the full propagator matrix is obtained by summing all chain transitions $a_0^0(980)\rightarrow f_0(980)\rightarrow\cdot\cdot\cdot\rightarrow a_0^0(980)$ and $f_0(980)\rightarrow a_0^0(980) \rightarrow\cdot\cdot\cdot\rightarrow f_0(980)$, respectively. The result is  \cite{Sekihara:2014qxa}:
\begin{equation}\label{GG}
\begin{split}
\left(
\begin{array}{cccc}
P_{a_0}(s)\quad P_{a_0f_0}(s)\\
 P_{f_0 a_0}(s)\quad P_{f_0}(s)\\
\end{array}
\right)=\frac{1}{D_{f_0}(s)D_{a_0}(s)-|D_{a_{0}f_{0}}(s)|^2}\left(
\begin{array}{cccc}
D_{a_0}(s)\quad D_{a_{0}f_{0}}(s)\\
D_{a_{0}f_{0}}(s)\quad D_{f_0}(s)\\
\end{array}
\right),
\end{split}
\end{equation}
where $P_{a_0}(s)$ and $P_{f_0}(s)$ represent the propagators of $a_0^0(980)$ and $f_0(980)$, respectively. The terms $P_{a_0f_0}(s)$, $P_{f_0 a_0}(s)$, and $D_{a_{0}f_{0}}(s)$ emerge as a consequence of the $a_0^0(980)-f_0(980)$ mixing effect. The individual propagators take the form:
\begin{equation}\label{Pa0f0}
	\begin{split}
	P_{a_{0}/f_{0}}(s)=\frac{D_{f_{0}/a_{0}(s)}}{D_{a_{0}/f_{0}(s)}(s)D_{f_{0}/a_{0}(s)}(s)-D_{a_{0}f_{0}}(s)^{2}}.
	\end{split}
\end{equation}

Meanwhile, $D_{a_0}(s)$ and $D_{f_0}(s)$ denote the denominators of the propagators for $a_0(980)$ and $f_0(980)$ in the absence of the $a_0^0(980)-f_0(980)$ mixing effect. These denominators can be expressed using the Flatt$\acute{\mathrm{e}}$ parametrization as follows:
\begin{equation}\label{DaDf}
\begin{split}
D_{a_0}(s)&=m_{a_0}^2-s-i\sqrt{s}[\Gamma_{\eta\pi}^{a_0}(s)+\Gamma_{K\bar{K}}^{a_0}(s)],\\
D_{f_0}(s)&=m_{f_0}^2-s-i\sqrt{s}[\Gamma_{\pi\pi}^{f_0}(s)+\Gamma_{K\bar{K}}^{f_0}(s)],\\
\end{split}
\end{equation}
where $m_{a_0}$ and $m_{f_0}$ are the masses of the $a_0(980)$ and $f_0(980)$ mesons, with the decay width $\Gamma^a_{bc}$ can being presented as:

\begin{equation}\label{Tabc}
\begin{split}
\Gamma_{bc}^a(s)=\frac{g_{abc}^2}{16\pi\sqrt{s}}\rho_{bc}(s)\quad \mathrm{with} \quad \rho_{bc}(s)=\sqrt{[1-\frac{(m_b-m_c)^2}{s}][1+\frac{(m_b-m_c)^2}{s}]}.\\
\end{split}
\end{equation}

It has been demonstrated that the contribution arising from the amplitude of $a_0^0(980)-f_0(980)$ mixing is convergent. When only the contributions from $K\bar{K}$ loop diagrams are taken into account, the amplitude can be expressed as an expansion within the $K\bar{K}$ phase space \cite{Achasov:1979xc,Achasov:2017zhu},
\begin{equation}\label{a0f0}
\begin{split}
D_{a_{0}f_{0}}(s)_{K\bar{K}}&=\frac{g_{a_0K^+K^-}g_{f_0K^+K^-}}{16\pi}\bigg\{i\bigg[\rho_{K^+K^-}(s)-\rho_{K^0\bar{K}^0}(s)\bigg]
-\mathcal{O}(\rho_{K^+K^-}^2(s)-\rho_{K^0\bar{K}^0}^2(s))\bigg\},\\
\end{split}
\end{equation}
where $g_{a_0K^+K^-}$ and $g_{f_0K^+K^-}$ are the effective coupling constants. Since the mixing mainly comes from the $K\bar{K}$ loops,
we can adopt $D_{a_{0}f_{0}} (s)\approx D_{a_{0}f_{0}}(s)_{K\bar{K}}$.

\subsection{$a_0^0(980)-f_0(980)$ mixing intensity}
There exist two types of reactions that can be utilized to investigate the $a_0^0(980)-f_0(980)$ mixing: $X\to Y f_0(980) \to Y a_0^0(980) \to Y \pi^{0}\eta$ and $X \to Y a_0^0(980) \to Y f_0(980)
\to Y \pi\pi$. The two processes correspond to two different types of mixing, namely $f_0(980) \to a_0^0(980)$ and $a_0^0(980) \to f_0(980)$.

\begin{figure}[ht]
\centerline{\includegraphics[width=0.8\textwidth]{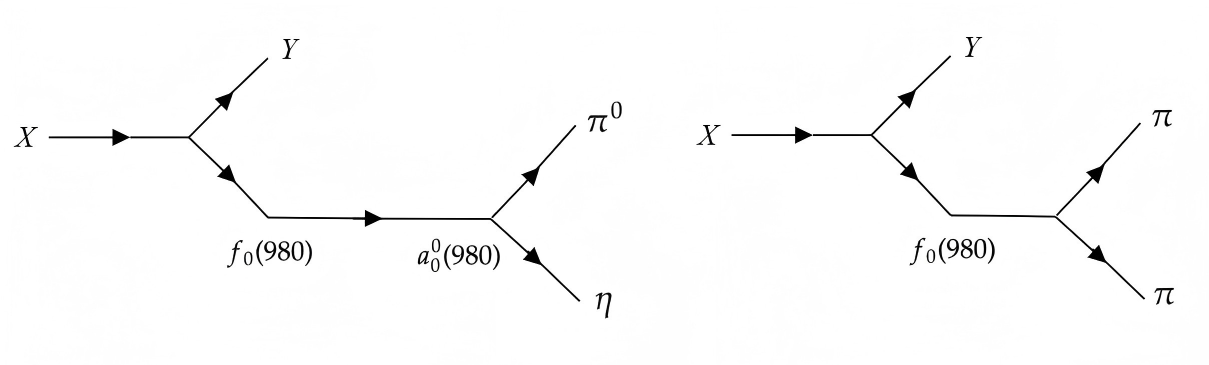}}
\caption{The Feynman diagram for $X\to Y f_0(980) \to Y a_0^0(980) \to Y \pi^{0}\eta$ (left) and $X\to Y f_0(980) \to Y \pi\pi$ (right) \cite{Wu:2008hx}.}
\label{faMixing}
\end{figure}

As shown on the left of Fig.~\ref{faMixing}, leaving only the contribution from $f_0(980) \to a_0^0(980)$, it is necessary to eliminate the influence caused by different $X$ and $Y$ particles,  and adding the contribution from the right of Fig.~\ref{faMixing}. Based on the two Feynman diagrams, the mixing intensity $\xi_{fa}$ for the transition $f_0(980) \to a_0^0(980)$ can be defined as \cite{Wu:2008hx}:
\begin{equation}
\begin{split}
    \xi_{fa}(s)&=\frac{d\Gamma_{X \to Y f_0(980) \to Y a_0^0(980) \to Y
\pi^{0}\eta}(s)}{d\Gamma_{X \to Y f_0(980) \to Y \pi\pi}(s)}\\
&=\frac{|D_{a_{0}f_{0}}|^{2}\Gamma^{a}_{\pi\eta}}{|D_{a_0}|^{2}\Gamma^{f}_{\pi\pi}}\label{mixing_intensity fa},
\end{split}
\end{equation}
where $s$ is the invariant mass squared of two mesons in final
state.

\begin{figure}[ht]
\centerline{\includegraphics[width=0.8\textwidth]{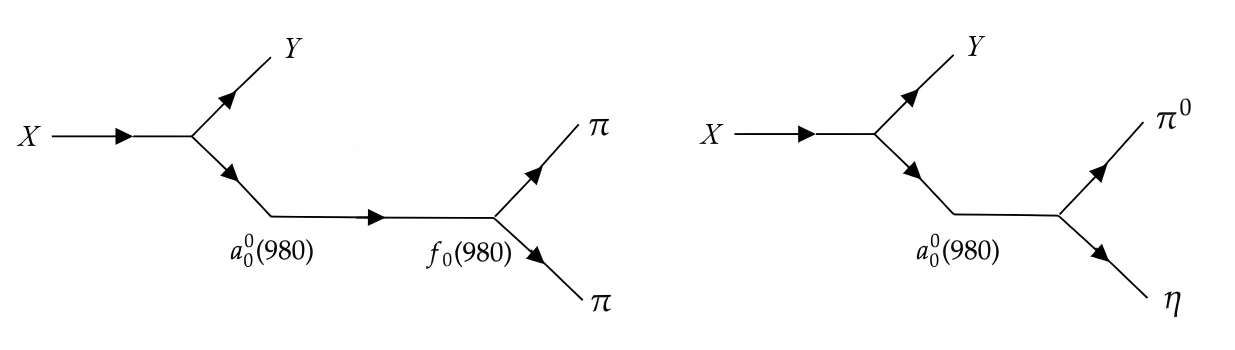}}
\caption{The Feynman diagram for $X \to Y a_0^0(980) \to Y f_0(980) \to Y \pi\pi$ (left) and $X \to Y a_0^0(980) \to Y \pi^{0}\eta$ (right).}
\label{afMixing}
\end{figure}

Similarly, for the reaction $X \to Y a_0^0(980) \to Y f_0(980) \to Y \pi\pi$ in the Fig.~\ref{afMixing}, the mixing intensity $\xi_{af}$ for the transition $a_0^0(980) \to f_0(980)$ is defined as follow \cite{Wu:2008hx}:
\begin{equation}
\begin{split}
    \xi_{af}(s)&=\frac{d\Gamma_{X \to Y a_0^0(980) \to Y f_0(980) \to Y
\pi\pi}(s)}{d\Gamma_{X \to Y a_0^0(980) \to Y \pi^{0}\eta}(s)}\\
&=\frac{|D_{a_{0}f_{0}}|^{2}\Gamma^{f}_{\pi\pi}}{|D_{f_0}|^{2}\Gamma^{a}_{\pi\eta}}\label{mixing_intensity af}.
\end{split}
\end{equation}

However, the physical quantities that can be directly measured by experiments are the integrated mixing intensities, which are defined as \cite{Wang:2016wpc}: 
\begin{equation}
    \overline  \xi_{fa}=\int_{s'_{\text{min}}}^{s'_{\text{max}}} ds \frac{\sqrt{s} |D_{a_{0}f_{0}}(s)|^2}{  |D_{f_0}(s) D_{a0}(s) |^2} \Gamma^a_{\pi\eta}(s) \bigg/ \int_{s_{\text{min}}}^{s_{\text{max}}} ds \frac{\sqrt{s}}{ |D_{f_0}(s)|^2} \Gamma^f_{\pi\pi}(s),  \label{eq:averaged_mixing_intensity fa}
\end{equation}

and

\begin{equation}
    \overline  \xi_{af}=\int_{s_{\text{min}}}^{s_{\text{max}}} ds \frac{\sqrt{s} |D_{a_{0}f_{0}}(s)|^2}{  |D_{f_0}(s) D_{a_0}(s) |^2} \Gamma^f_{\pi\pi} (s) \bigg/ \int_{s'_{\text{min}}}^{s'_{\text{max}}} ds \frac{\sqrt{s}}{ |D_{a_0}(s)|^2} \Gamma^a_{\pi\eta}(s),  \label{eq:averaged_mixing_intensity af}
\end{equation}
where the terms $s^{(')}_{\text{min}}$ and $s^{(')}_{\text{max}}$ designate the minimum and maximum values of the invariant mass cuts, respectively.

\section{Decay amplitudes and CP asymmetries}
\label{Amp Acp}

Following the treatment of 
$B$ decays in Refs. \cite{Wang:2019tqt,Qi:2023pls}, the total decay amplitude for the process $D^\pm \rightarrow f_0(980)[a_0^0(980)]\pi^\pm \rightarrow \pi^+\pi^-\pi^\pm$ incorporating $a_0^0(980) - f_0(980)$ mixing can be expressed as:
\begin{equation}\label{M3}
\begin{split}
\mathcal{M}(D^+\rightarrow \pi^+\pi^-\pi^+)=\frac{g_{f_0\pi\pi}}{D_{f_0}}\mathcal{M}(D^+\rightarrow f_0\pi^+)
+\frac{g_{f_0\pi\pi}D_{a_{0}f_{0}}}{D_{a_0}D_{f_0}-D_{a_{0}f_{0}}^2}\mathcal{M}(D^+\rightarrow a_0^0\pi^+),\\
\end{split}
\end{equation}
where $g_{f_0\pi\pi}$ is the coupling constant, $\mathcal{M}(D^+\rightarrow f_0\pi^+)$ and $\mathcal{M}(D^+\rightarrow a_0^0\pi^+)$ are the decay amplitudes of the processes $D^+\to f_0(980)\pi^+$ and $D^+\to a_0(980)\pi^+$, respectively.

For the two weak decays $D^+\to f_0(980)\pi^+$ and $D^+\to a_0(980)\pi^+$, the corresponding effective Hamiltonian can be expressed as: \cite{Buchalla:1995vs,Guo:2000pt}
\begin{equation}
H_{\Delta C=1}=\frac{G_F}{\sqrt{2}}\left\{\left[\sum_{q=d,s}V_{uq}V_{cq}^{\ast}\left(c_1 O^q_1 + c_2 O^q_2\right)\right]-V_{ub}V_{cb}^{\ast}\sum_{i=3}^{6} c_i O_i\right\}+\text{h.c.},
\end{equation}
where $G_{F}$ is the Fermi constant, $V_{q_1q_2}$ ($q_1$ and $q_2$ represent quarks) are the CKM matrix elements, $c_{i}$ ($i=1,\cdots,6$) are the Wilson coefficient, and the four-quark operators $O_i$ are
\begin{eqnarray}
O_{1}^{q}& = &\bar{u}_{\alpha} \gamma_{\mu}(1-\gamma{_5})q_{\beta}\bar{q}_{\beta} \gamma^{\mu}(1-\gamma{_5})
c_{\alpha}\ , \nonumber \\
 O_{2}^{q}& = &\bar{u} \gamma_{\mu}(1-\gamma{_5})q\bar{q} \gamma^{\mu}(1-\gamma{_5})c\ ,  \nonumber \\
O_{3}& =& \bar{u} \gamma_{\mu}(1-\gamma{_5})c \sum_{q\prime}\bar{q}^{\prime}\gamma^{\mu}(1-\gamma{_5})
q^{\prime}\ ,\nonumber \\
O_{4}& =&\bar{u}_{\alpha} \gamma_{\mu}(1-\gamma{_5})c_{\beta}
\sum_{q\prime}\bar{q}^{\prime}_{\beta}\gamma^{\mu}(1-\gamma{_5})q^{\prime}_{\alpha}\ , \nonumber \\
O_{5}& =&\bar{u} \gamma_{\mu}(1-\gamma{_5})c \sum_{q'}\bar{q}^
{\prime}\gamma^{\mu}(1+\gamma{_5})q^{\prime}\ , \nonumber \\
O_{6}& =&\bar{u}_{\alpha} \gamma_{\mu}(1-\gamma{_5})c_{\beta}
\sum_{q'}\bar{q}^{\prime}_{\beta}\gamma^{\mu}(1+\gamma{_5})q^{\prime}_{\alpha}\ , 
\end{eqnarray}
with $O_{1-2}^{q}$ being tree operators, $O_{3-6}$  QCD penguin operators, $\alpha$, $\beta$ color indices, and the sum over $q'$ running over all  light flavour quarks.

In our calculations, there often refer to the decay constants and form factors generally. The decay constants of the scalar meson $S$ and pseudoscalar meson $P$ are defined as, respectively \cite{Cheng:2022vbw}, 
\begin{equation}\label{a}
\langle S(p)|\bar{q}_2\gamma_\mu q_1|0\rangle=\bar{f}_Sp_\mu,
~~~~~\langle P(p)|\bar{q}_2\gamma_\mu\gamma_5 q_1|0\rangle=-if_Pp_\mu.
\end{equation}

The form factors of $D\rightarrow S,P$ are defined by \cite{Bauer:1986bm}:
\begin{equation}\label{bv}
\begin{split}
\langle P(p')|\hat{V}_\mu|D(p)\rangle&=\bigg(p_\mu-\frac{m_D^2-m_P^2}{q^2}q_\mu\bigg) F_1^{DP}(q^2)+\frac{m_D^2-m_P^2}{q^2}q_\mu F_0^{DP}(q^2),\\
\langle S(p')|\hat{A}_\mu|D(p)\rangle&=-i\bigg[\bigg(P_\mu-\frac{m_D^2-m_S^2}{q^2}q_\mu\bigg)F_1^{DS}(q^2)+\frac{m_D^2-m_S^2}{q^2}q_\mu F_0^{DS}(q^2)\bigg],\\
 \end{split}
\end{equation}
where $P_\mu=(p+p')_\mu$, $q_\mu=(p-p')_\mu$, $\hat{V}_\mu$ and $\hat{A}_\mu$ are the weak vector and axial-vector currents, respectively. $\hat{V}_\mu=\bar{q}_f\gamma_\mu c, \hat{A}_\mu=\bar{q}_f\gamma_\mu \gamma_5c$, with $q_f$ being the quarks generated from the decay of $c$ quark. As for the $F(q^2)$, we use the 3-parameter parametrization:
\begin{equation}\label{a}
F(q^2)=\frac{F(0)}{1-a(q^2/m_D^2)+b(q^2/m_D^2)^2},
\end{equation}
for $D\rightarrow SP$ decays, the relevant form factors will be adopted the $F_0^{DS}(q^2)$ and $F_0^{DP}(q^2)$ in this work.

The factorization amplitudes for the $D\rightarrow SP$ read \cite{Cheng:2022vbw}:
\begin{equation}\label{a}
\begin{split}
X^{(DS,P)}&=\langle P(q)|(V-A)_\mu|0\rangle \langle S(p)|(V-A)^\mu|D(p_D)\rangle,\\
X^{(DP,S)}&=\langle S(q)|(V-A)_\mu|0\rangle \langle P(p)|(V-A)^\mu|D(p_D)\rangle.\\
\end{split}
\end{equation}

In the two-quark model with ideal  mixing for $f_0(980)$ and $\sigma(600)$,  $f_0(980)$ is characterized as a pure $s\bar{s}$ state, while $\sigma(600)$ is identified as an $n\bar{n}$ state, where $n\bar{n} \equiv (\bar{u}u + \bar{d}d)/\sqrt{2}$. Nevertheless, experimental evidence suggests that $f_0(980)$ does not consist solely of an $s\bar{s}$ configuration. A notable example is the observation of $\Gamma(J/\psi \to f_0\omega) \approx \frac{1}{2}\Gamma(J/\psi \to f_0\phi)$ \cite{ParticleDataGroup:2024cfk}, which unequivocally demonstrates the presence of both non-strange and strange quark components within $f_0(980)$. Consequently, it is imperative that the isoscalar states $\sigma(600)$ and $f_0(980)$ exhibit a mixing phenomenon, like
\begin{equation}
    \begin{split}
        |f_0(980)\rangle &= |s\bar{s}\rangle \cos \theta + |n\bar{n}\rangle \sin \theta,\\
        |\sigma(600)\rangle &= -|s\bar{s}\rangle \sin \theta + |n\bar{n}\rangle \cos \theta,
    \end{split}
\end{equation}
where $\theta$ is the mixing angle of
the $f_0(980)$ and $\sigma(600)$.

Within the naive factorization approach 
\cite{Bjorken:1988kk}, the decay amplitudes of $D^+\rightarrow f_0\pi^+$ and $D^+\rightarrow a_0^0\pi^+$ are
\begin{equation}\label{naive amplitude}
\begin{split}
\mathcal{M}(D^+\rightarrow f_0\pi^+)&=\frac{G_F}{\sqrt{2}}\Bigg\{ V_{ud}V_{cd}^\ast\left(\frac{\alpha}{\sqrt{2}}a_1\bar{f}^d_{f_0}X^{D\pi}_{f_0}-a_2 Y^{Df_0}_{\pi}\right)+V_{us}V_{cs}^\ast \beta a_1\bar{f}^s_{f_0}X^{D\pi}_{f_0}\\
&+V_{ub}V_{cb}^\ast\bigg[-a_3\Big(\frac{\alpha}{\sqrt{2}}\bar{f}^u_{f_0}X^{D\pi}_{f_0}+\frac{\alpha}{\sqrt{2}}\bar{f}^d_{f_0}X^{D\pi}_{f_0}+\beta\bar{f}^s_{f_0}X^{D\pi}_{f_0}\Big)+a_4Y^{Df_0}_{\pi}\\
&-a_5\Big(\frac{\alpha}{\sqrt{2}}\bar{f}^u_{f_0}X^{D\pi}_{f_0}+\frac{\alpha}{\sqrt{2}}\bar{f}^d_{f_0}X^{D\pi}_{f_0}+\beta\bar{f}^s_{f_0}X^{D\pi}_{f_0}\Big)-\frac{2m_\pi^2a_6}{(m_c+m_d)(m_u+m_d)}Y^{Df_0}_{\pi}\bigg]\Bigg\},\\
\mathcal{M}(D^+\rightarrow a_0^0\pi^+)&=\frac{G_F}{\sqrt{2}}\Bigg\{ V_{ud}V_{cd}^\ast\left(-\frac{1}{\sqrt{2}}a_1\bar{f}_{a_0}X^{D\pi}_{a_0}-a_2Y^{Da_0}_{\pi}\right)\\
&+V_{ub}V_{cb}^\ast\bigg[a_4Y^{Da_0}_{\pi}-\frac{2m_\pi^2a_6}{(m_c+m_d)(m_u+m_d)}Y^{Da_0}_{\pi}\bigg]\Bigg\},
\end{split}
\end{equation}
respectively, where $(\alpha,\beta)=(\sin\theta,\cos\theta)$, $f_\pi$, $\bar{f}_{f_0}$ and $\bar{f}_{a_0}$ are the decay constants of the $\pi$, $f_0(980)$ and $a_0^0(980)$ mesons, respectively, $F_0^{D\pi}$, $F_0^{Df_0}$ and $F_0^{Da_0}$ are the corresponding transition form factors, and $a_i$ are built up from the Wilson coefficients $c_{i}$ with the form $a_i=c_i+c_{i\pm1}/N_c$. For compactness, we can use the form: $X^{D\pi}_{f_0/a_0}=(m^2_D-m_\pi^2)F_0^{D\pi}(m_{f_0/a_0}^2)$ and $Y^{Df_0/a_0}_{\pi}=f_\pi(m^2_D-m_{f_0/a_0}^2)F_0^{Df_0/a_0}(m_{\pi}^2)$.

For the $CP$-conjugate process $D^\pm \to \pi^+\pi^-\pi^\pm$, the direct $CP$ asymmetry can be expressed in the following form:
\begin{equation}\label{ACP}
\begin{split}
A_{CP}=&\frac{\mid\mathcal{M}\mid^2-\mid\mathcal{\bar{M}}\mid^2}{\mid\mathcal{M}\mid^2+\mid\mathcal{\bar{M}}\mid^2}.
\end{split}
\end{equation}

\section{Numerical results}
\label{Num res}
From Eqs. (\ref{mixing_intensity fa}) to (\ref{eq:averaged_mixing_intensity af}), we can find that the $\xi_{fa/af}$ and $\overline\xi_{fa/af}$ depend on the $g_{f_0\pi^0\pi^0}$, $g_{f_0K^+K^-}$, $g_{a_0\pi^0\eta}$, $g_{a_0K^+K^-}$, $m_{a_0}$ and $m_{f_0}$. 
 These parameters, compiled from a range of theoretical models and experimental analyses \cite{Achasov:1987ts,Achasov:1997ih,Weinstein:1983gd,Weinstein:1990gu,Achasov:2000ym,Achasov:2000ku,KLOE:2002kzf,KLOE:2002deh,E852:1996san,Zou:1993az,Bugg:1994mg,BES:2004twe}, are listed in Table ~\ref{tab:masses&coupling}. The resulting predictions for 
$\xi_{fa}$ and $\xi_{af}$ evaluated at $\sqrt{s}=991.3$ MeV, together with the integrated quantities $\overline\xi_{fa}$ and $\overline\xi_{af}$, are presented in Table~\ref{tab:mixing_values}, and are consistent with Refs.~\cite{Wu:2008hx,Wang:2016wpc}. The differential mixing intensity satisfies 
$\xi_{fa}>\xi_{af}$ from different models, whereas the integrated quantities exhibit the opposite ordering, 
$\bar{\xi}_{fa}<\bar{\xi}_{af}$, making 
$\bar{\xi}_{af}$ more favorable observable for experimental measurement. For the decay 
$D^\pm \rightarrow a_0^0(980)\pi^\pm \rightarrow f_0(980)\pi^\pm \rightarrow \pi^+\pi^-\pi^\pm$, we focus on the $\bar{\xi}_{af}$. The majority of theoretical predictions place $\bar{\xi}_{af}$
in the percent range; models B and C and experiment E yield particularly large values. This also implies that the $a_0^0(980)-f_0(980)$ mixing mechanism may have a non-negligible impact on the $CP$ violation. The mesons $f_0(980)$ and $a_0^0(980)$ can be described within various theoretical frameworks, including as conventional $q\bar{q}$ states, $qq\bar{q}\bar{q}$ multiquark configurations, meson-meson bound states, or even scalar glueballs. If these scalar mesons are interpreted as four-quark states, the decay process necessitates the production of an additional quark-antiquark pair compared to the two-quark scenario. Consequently, it is anticipated that the decay amplitude in the four-quark picture would be suppressed relative to that in the two-quark picture when a light scalar meson is involved. For this reason, we adopt the assumption that the $q\bar{q}$ structure is dominant in our analysis.

\begin{table}[h]
\setlength{\tabcolsep}{8pt}
\caption{Meson masses (in units of MeV) and coupling canstants (in units of GeV) from various models or determined by experimental measurements.}\label{tab:masses&coupling}
\begin{tabular}{cccccccccc}
\hline  No. & model/experiment & $m_{a_0}$ & $g_{a_{0}\pi\eta}$ &
$g_{a_{0}K^{+}K^{-}}$ & $m_{f_0}$ & $g_{f_{0}\pi^{0}\pi^{0}}$ & $g_{f_{0}K^{+}K^{-}}$  \\
\hline
  A & $q\bar{q}$ model \cite{Achasov:1987ts} & 983 & 2.03 & 1.27 & 975 & 0.64 & 1.80  \\

  B & $q^{2}\bar{q}^{2}$ model \cite{Achasov:1987ts}  & 983 & 4.57 & 5.37 & 975 & 1.90 & 5.37 \\

  C & $K\bar{K}$ model \cite{Achasov:1997ih,Weinstein:1983gd,Weinstein:1990gu}  & 980 & 1.74 & 2.74 & 980 & 0.65 & 2.74  \\

  D & $q\bar{q}g$ model \cite{Wu:2008hx} & 980  & 2.52 & 1.97 & 975 & 1.54 & 1.70  \\

 E & SND \cite{Achasov:2000ym,Achasov:2000ku} & 995 & 3.11 & 4.20 & 969.8 & 1.84 & 5.57 \\

 F & KLOE \cite{KLOE:2002kzf,KLOE:2002deh} & 984.8  & 3.02 & 2.24 & 973 & 2.09 & 5.92 \\

 G & BNL~\cite{E852:1996san,Zou:1993az} & 1001  & 2.47 & 1.67 & 953.5 & 1.36 & 3.26 \\
 
 H & CB~\cite{Bugg:1994mg,BES:2004twe} & 999 & 3.33 & 2.54  & 965 & 1.66  & 4.18 \\
\hline
\end{tabular}
\end{table}

\begin{table}[h]
\setlength{\tabcolsep}{8pt}
\caption{The mixing intensity $\xi_{fa}(s)$ and $\xi_{af}(s)$ are evaluated at $\sqrt{s}=991.3$ MeV, which is at the center of the $K^+K^-$ and $K^0\bar K^0$ threshold. The integrated  mixing intensity $\bar \xi_{fa}$ and $\bar \xi_{af}$(in unit  of $\%$) are evaluated by Eqs.~(\ref{eq:averaged_mixing_intensity fa}) and (\ref{eq:averaged_mixing_intensity af}) with the kinematics in Eqs.~(\ref{eq:pieta_cuts}) and (\ref{eq:pipi_cuts}).  }\label{tab:mixing_values}
\begin{tabular}{c|ccc|cc|cc|c}
\hline  
\multirow{2}{*}{No.} & \multicolumn{3}{c|}{$\xi_{fa}(\%)$} & \multicolumn{2}{c|}{$\xi_{af}(\%)$} & \multicolumn{2}{c|}{$\overline \xi_{fa}(\%)$} & $\overline \xi_{af}(\%)$ \\
\cline{2-9}
   & This work & Ref. \cite{Wu:2008hx} & Ref. \cite{Wang:2016wpc} & This work & Ref. \cite{Wu:2008hx} & This work & Ref. \cite{Wang:2016wpc} & This work \\
   \hline
  A & 2.24 & 2.30 & 2.20 & 0.94 & 1.00 & 0.04 & 0.90 & 2.19 \\

  B & 6.57 & 6.80 & 6.50 & 5.99 & 6.20 & 0.88 & 1.80 & 10.26 \\

  C & 20.47 & 21.00 & 20.10 & 14.85 & 15.00& 7.77 &11.10 & 27.21 \\

  D & 0.52 & 0.50 & ... & 0.57 & 0.60& 0.06 &...& 1.08 \\

  E & 8.61 & 8.80 & 8.50 & 8.60 & 8.90& 1.68 & 2.60 & 14.32 \\

  F & 3.24 & 3.40 & 3.20 & 2.39 & 2.50 & 0.57  &0.80 &  4.30 \\

  G & 1.83 & 1.90 & 1.80 & 1.35 & 1.40 & 0.21 & 0.50 & 2.55 \\
 
  H & 2.61 & 2.70 & 2.60 & 2.17 & 2.30 & 0.40 & 0.70 & 3.97 \\
\hline
\end{tabular}
\end{table}

\begin{figure}[h]
\centering
\subfigure[]{
\includegraphics[width=0.45\columnwidth]{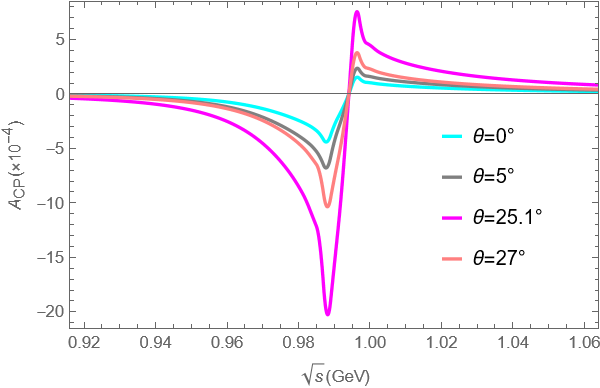}
}
\,
\subfigure[]{
\includegraphics[width=0.456\columnwidth]{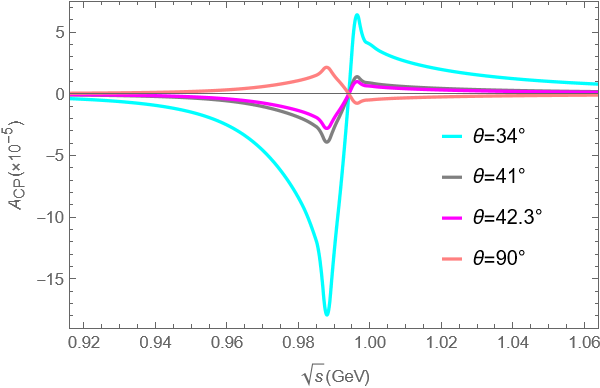}
}
\caption{The differential $CP$-violating asymmetry as a function of $\sqrt{s}$ in the decay $D^\pm \rightarrow f_0(980)[a_0^0(980)]\pi^\pm \rightarrow \pi^+\pi^-\pi^\pm$.}
\label{fig1}
\end{figure}

Substituting Eq. (\ref{naive amplitude}) into Eq. (\ref{M3}), one can get the total amplitude of the $D^\pm \rightarrow f_0(980)[a_0^0(980)]\pi^\pm \rightarrow \pi^+\pi^-\pi^\pm$ decay with the $a_0^0(980)-f_0(980)$ mixing mechanism. Fig. ~\ref{fig1} shows the differential $CP$ asymmetry as a function of  $\sqrt{s}$
  for several values of the mixing angle 
$\theta$. The asymmetry varies rapidly and changes sign in the vicinity of the 
$f_0(980)$ resonance. For $\theta=0^\circ$ ($|f_0(980)\rangle = |s\bar{s}\rangle$),  $A_{CP}$  ranges from $-4.40 \times 10^{-4}$ to $1.52 \times 10^{-4}$, for $\theta=90^\circ$ ($|f_0(980)\rangle = |n\bar{n}\rangle = \frac{1}{\sqrt{2}}|u\bar{u} + d\bar{d}\rangle$), it is substantially reduced, ranging from $-0.07 \times 10^{-4}$ to $0.22 \times 10^{-4}$. In reality, the $f_0(980)$ is a mixed state with $n\bar{n}$ and $s\bar{s}$. Taking the central value provided in \cite{Cheng:2005nb}, we have examined several examples, which demonstrate that the range of $A_{CP}$ is highly sensitive to the mixing angle when both 
$n\bar{n}$ and $s\bar{s}$ components are present. The ranges of $A_{CP}$ corresponding to these central angles are summarized in Table~\ref{tab:3}. From these results, we can clearly see that the $a_0^0(980)-f_0(980)$ mixing mechanism can relatively increase the $CP$ violation, especially for $\theta=25.1^\circ$ and $\theta=27^\circ$ conditions, reaching $10^{-4}$ to $10^{-3}$. With the improvement of experimental precision, constraining the range of the mixing angle can also help us further confirm the range of $CP$ violation.

\begin{table}[h]
\setlength{\tabcolsep}{15pt}
\caption{Ranges of the direct $CP$ asymmetry $A_{CP}$ for different central mixing angles $\theta$, where $R\equiv g_{f_{0}K^{+}K^{-}}^{2}/g_{f_{0}\pi^{+}\pi^{-}}^{2}$  
measures the ratio of the $f_0(980)$ coupling to $K^+K^-$ and $\pi^+\pi^-$.}\label{tab:3}
\begin{tabular}{ccccc}
\hline   Experimental implications & Mixing angle & $A_{CP}$ & Reference \\
\hline
   pure $s\bar{s}$ state & $\theta=0^{\circ}$ & $-4.40 \times 10^{-4}$ to $1.52 \times 10^{-4}$ & ... \\

  $\phi\to f_0\gamma,f_0\to\gamma\gamma$ & $\theta=5^{\circ}$ & $-6.78 \times 10^{-4}$ to $2.35 \times 10^{-4}$ & \cite{Anisovich:2001zp} \\
  
  $R=4.03\pm0.14$ & $\theta=25.1^{\circ}$ & $-20.21 \times 10^{-4}$ to $7.55 \times 10^{-4}$ & \cite{KLOE:2002deh} \\
  
  QCD sum rules and $f_0$ data & $\theta=27^{\circ}$ & $-10.29 \times 10^{-4}$ to $3.76 \times 10^{-4}$ & \cite{Cheng:2002ai} \\
  
  $J/\psi\to f_0\phi,f_0\omega$ & $\theta=34^{\circ}$ & $-1.79 \times 10^{-4}$ to $0.63 \times 10^{-4}$ & \cite{Achasov:2002xd} \\
  
  QCD sum rules and $a_0$ data & $\theta=41^{\circ}$ & $-0.39 \times 10^{-4}$ to $0.14 \times 10^{-4}$ & \cite{Cheng:2002ai} \\
  
  $R=1.63\pm0.46$ & $\theta=42.3^{\circ}$ & $-0.28 \times 10^{-4}$ to $0.10 \times 10^{-4}$ & \cite{Fleischer:1999hp} \\

   pure $n\bar{n}$ state & $\theta=90^{\circ}$ & $-0.07 \times 10^{-4}$ to $0.22 \times 10^{-4}$ & ... \\
  
\hline
\end{tabular}
\end{table}

\section{SUMMARY AND DISCUSSION}
\label{Sum Dis}

We have studied the direct $CP$ violation in 
$D^\pm \to \pi^\pm \pi^+ \pi^-$ decay incorporating the 
 $a_0^0(980)-f_0(980)$ mixing mechanism within the naive factorization approach. The integrated mixing intensity $\bar{\xi}_{af}$
is found to be of order percent for several input parameter sets, confirming that this isospin-breaking effect is phenomenologically significant. Applying the mechanism to the $CP$ asymmetry calculation, we find that the differential $CP$ violation is enhanced to $\mathcal{O}$($10^{-4}$-$10^{-3}$) depending on the 
$f_0(980)$ and $\sigma(600)$ mixing angle 
$\theta$. Since different experimental and theoretical determinations of 
$\theta$ are not yet mutually consistent, we present results across the full range of reported central values rather than adopting a single value. The analysis demonstrates that the 
$a_0^0(980)-f_0(980)$ mixing mechanism should be systematically considered in amplitude analyses of 
$D$ or $B$ meson three-body decays.

\acknowledgments
This work was supported by National
Natural Science Foundation of China under Grants Nos.  12405115, 12105149, 11805153 and 12275024. 

\appendix
\section{Theoretical input parameters}
\label{A}

We adopt the $s^{(')}_{min}$ and $s^{(')}_{max}$ from the Ref. \cite{Wang:2016wpc}:
\begin{eqnarray}
 s'_{min}= [(991.3-4){\rm MeV}]^2,\;\;\;  s'_{max}= [(991.3+4){\rm MeV}]^2, \label{eq:pieta_cuts}
\end{eqnarray}
and
\begin{eqnarray}
 s_{min} = [900{\rm MeV}]^2, \;\;\;
 s_{max} = [1000{\rm MeV}]^2.  \label{eq:pipi_cuts}
\end{eqnarray}

We use the Wolfenstein parameterization for the CKM matrix elements, which up to the order of $\lambda^8$ , can be expressed as \cite{Charles:2004jd}:
\begin{eqnarray}
V_{ud}&=&1-\frac{\lambda^2}{2}-\frac{\lambda^4}{8}-\frac{\lambda^6}{16}[1+8A^2(\rho^2+\eta^2)]-\frac{\lambda^8}{128}[5-32A^2(\rho^2+\eta^2)],\nonumber\\
V_{cd}&=&-\lambda+\frac{\lambda^5}{2}A^2[1-2(\rho+i\eta)]+\frac{\lambda^7}{2}A^2(\rho+i\eta),\nonumber\\
V_{us} &=&\lambda - \frac{1}{2}A^2\lambda^7(\rho^2+\eta^2),\nonumber\\
V_{cs} &=&1-\frac{1}{2}\lambda^2-\frac{1}{8}\lambda^4(1+4A^2)- \frac{1}{16}\lambda^6\left(1 - 4A^2 + 16A^2(\rho+i\eta)\right)-\frac{1}{128}\lambda^8\left(5-8A^2+16A^4\right), \nonumber\\
V_{ub}&=&\lambda^3A(\rho-i\eta),\nonumber\\
V_{cb}&=&A\lambda^2-\frac{\lambda^8}{2}A^3(\rho^2+\eta^2),
\end{eqnarray}
with $A$, $\rho$, $\eta$, and $\lambda$ being the Wolfenstein parameters. We use
the results in Ref. \cite{ParticleDataGroup:2018ovx}:
\begin{equation}\label{CKM}
\begin{split}
&\lambda=0.22465\pm0.00039,\quad A=0.832\pm0.009,\\
&\bar{\rho}=0.139\pm0.016, \quad \bar{\eta}=0.346\pm0.010, \\
\end{split}
\end{equation}
where
\begin{equation}\label{rhoyita}
\begin{split}
\bar{\rho}=\rho(1-\frac{\lambda^2}{2}), \quad \bar{\eta}=\eta(1-\frac{\lambda^2}{2}). \\
\end{split}
\end{equation}

The effective Wilson coefficients used in our calculations are taken from Ref. \cite{Guo:1998eg}:
\begin{equation}\label{C}
\begin{split}
&c_1=-0.6941,\quad c_2=1.3777,\quad c_3=0.0652,\\
&c_4=-0.0627,\quad c_5=0.0206,\quad c_6=-0.1355.\\
\end{split}
\end{equation}

For the masses appeared in $D$ decays, we use the following values \cite{Cheng:2005nb,ParticleDataGroup:2018ovx} (in units of $\mathrm{GeV}$):
\begin{equation}
\begin{split}
&m_u=0.0035,\quad m_d=0.0063,\quad m_s=0.119, \quad m_c=1.3,\\
&m_{\eta}=0.5475,\quad m_{\pi^\pm}=0.1396,\quad 
m_{K^\pm}=0.4937,\quad m_{K^0}=0.4977,\quad\\
&m_{D^\pm}=1.870,\quad m_{f_0(980)}=0.990,\quad m_{a_0^0(980)}=0.980,\\
\end{split}
\end{equation}

For the widths, we use \cite{ParticleDataGroup:2018ovx}(in units of GeV):
\begin{equation}
\Gamma_{f_0(980)}=0.074,\quad\Gamma_{a^0_0(980)}=0.092.
\end{equation}

As for the form factors, we use \cite{Cheng:2003sm,Soni:2020sgn}:
\begin{equation}
\begin{split}
F_0^{D\pi}(0)&=0.67, ~\text{with}~a=0.50, b=0.01,\\
F_0^{Df_0}(0)&=0.45, ~\text{with}~a=1.36, b=0.32,\\
F_0^{Da_0}(0)&=0.55, ~\text{with}~a=1.06, b=0.16.\\
\end{split}
\end{equation}

The following numerical values for the decay constants are used \cite{Cheng:2022vbw} (in units of $\mathrm{GeV}$):
\begin{equation}
\begin{split}
f_{\pi^\pm}=0.131,\quad
\bar{f}_{f_0(980)}^u= \bar{f}_{f_0(980)}^d=0.350\pm0.02,\\\bar{f}_{f_0(980)}^s=0.370\pm0.02, \quad\bar{f}_{a_0^0(980)}=0.365\pm0.02.
\end{split}
\end{equation}

\end{document}